\documentclass[useAMS,usenatbib,referee]{biom}

\usepackage{amsmath,amssymb}
\usepackage{setspace,subfig}
\usepackage{natbib}
\usepackage{relsize}

\usepackage{booktabs} 

\usepackage{caption} 
\usepackage{xfrac}
\usepackage{graphicx}

\usepackage{tikz}
\usepackage{times}
\usepackage{url}
\usetikzlibrary{arrows,shapes.arrows,shapes.geometric,
	shapes.multipartb,backgrounds,decorations.pathmorphing,positioning,fit,automata}
\tikzset{
	>=stealth',
	true/.style={
		rectangle,
		draw=black, very thick,
		text width=6.5em,
		minimum height=2em,
		text centered,
		fill=gray, opacity = 0.5},
	punkt/.style={
		rectangle,
		rounded corners,
		draw=black, very thick,
		text width=6.5em,
		minimum height=2em,
		text centered},
	est/.style={
		circle,
		draw=black, very thick,
		text centered},
	shade/.style={
		circle,
		draw=black, very thick, fill=gray!50,
		text centered},
	weight/.style={
		circle,
		draw=black, very thick,
		text width=6.5em,
		minimum height=2em,
		text centered},
	pil/.style={
		->,
		thick,
		shorten <=2pt,
		shorten >=2pt,},
	double/.style={
		<->,
		thick,
		shorten <=2pt,
		shorten >=2pt,},
	dash/.style={
		dashed,
		thick,
		shorten <=2pt,
		shorten >=2pt,},
	dashdouble/.style={
		<->,
		dashed,
		thick,
		shorten <=2pt,
		shorten >=2pt,}
}
\usepackage[T1]{fontenc}
\allowdisplaybreaks

\usepackage{xr}

\makeatletter
\newcommand*{\addFileDependency}[1]{
	\typeout{(#1)}
	\@addtofilelist{#1}
	\IfFileExists{#1}{}{\typeout{No file #1.}}
}
\makeatother

\newcommand*{\myexternaldocument}[1]{%
	\externaldocument{#1}%
	\addFileDependency{#1.tex}%
	\addFileDependency{#1.aux}%
}
\myexternaldocument{supple}

\usepackage{algorithm}
\usepackage{tikz}

\usepackage{soulpos}
\usepackage{multirow}
\definecolor{mygreen}{RGB}{144,241,47}

\usepackage[none]{hyphenat}

\usepackage[figuresright]{rotating}

\title[Rejoinder]{Rejoinder to discussions on ``Instrumental variable estimation of the causal hazard ratio''}

\author
{Linbo Wang\emailx{linbo.wang@utoronto.ca} \\
	Department of Statistical Sciences, University of Toronto, Toronto,
	Ontario M5S 3G3, Canada
	\and
	Eric Tchetgen Tchetgen\emailx{ett@wharton.upenn.edu} \\
	Department of Statistics, University of Pennsylvania, Philadelphia, Pennsylvania 19104, U.S.A.
	\and
	Torben Martinussen\emailx{tma@sund.ku.dk} \\
	Department of Biostatistics, University of Copenhagen, {\O}ster  Farimagsgade 5, 1014
	Copenhagen, Denmark
	\and
	Stijn Vansteelandt\emailx{stijn.vansteelandt@ugent.be} \\
	Department of Applied Mathematics, Computer Science and Statistics, Ghent University,
	Krijgslaan 281 (S9),\\ 9000 Ghent, Belgium}

\begin{document}
	
	
	
	\volume{63}
	\pubyear{2007}
	\artmonth{December}
	
	\doi{}
	
	\label{firstpage}
	
	
		
	
	
	\maketitle

	\section{Introduction}
	
	We thank the editors for the opportunity to publish our paper with discussions, and the discussants for their insightful and thought-provoking comments. We respond to their comments in the following order. In  Section \ref{sec:rr}, motivated by \cite{omalley2022discussion}, we discuss whether the risk difference is less likely to be homogeneous than other measures such as the odds ratio. In Section \ref{sec:identification}, we discuss the extensions of our identification approach to related causal parameters, including the cumulative distribution functions and the complier hazard ratio. In Section \ref{sec:specification}, we outline a specification test for the proportional hazard assumption in the IV context. In 
	Section \ref{sec:mr}, we comment on \cite{baer2022discussion}'s multiply robust estimator. We respond to the other comments in Section \ref{sec:others}.
	
	\section{Is the risk difference really less likely to be homogeneous?}
	\label{sec:rr}
	
	In the case of a randomized $Z$,  Assumption  5 in \cite{wang2022instrumental}
	implies no effect modification by $U$ for the effect of $Z$ on $X$ \emph{on the risk difference scale}. As pointed out by \cite{omalley2022discussion},  this typically implies effect modification on other scales such as the odds ratio. \cite{omalley2022discussion} argue that for a binary outcome, one would ``naturally expect''   the risk difference to be less likely to be homogeneous than other measures such as the odds ratio. They provide a specific example that if the probability $E[D\mid Z=0, X=x, U=u]$ was 0.95, then the risk difference $E[D\mid Z=1,X,U] - E[D\mid Z=0,X,U]$ is less likely to be homogeneous compared to if the baseline probability is near 0.5.
	
	We believe this represents a common confusion about whether a measure is likely to be homogeneous, and whether a measure is variation independent with a nuisance measure that is easier to interpret. The former is decided by nature, while the latter is specific to human beings. In particular, we argue that the conjecture provided in \cite{omalley2022discussion} reflects the variation dependence between the risk difference (RD) and the baseline risk, a parameter that is often used in practice. In comparison, the OR is variation independent of the baseline risk. There are, however, other parameters that are variation independent of the RD, such as the odds product (OP) introduced in \cite{richardson2017modeling}, defined as $$OP(X,U) = \dfrac{E[D\mid Z=1,X,U]E[D\mid Z=0,X,U]}{(1-E[D\mid Z=1,X,U])(1-E[D\mid Z=0,X,U])}.$$ Consider the following three statements, where $RD(X,U) = E[D\mid Z=1,X,U] - E[D\mid Z=0,X,U],$ and $OR(X,U) = \dfrac{E[D\mid Z=1,X,U](1-E[D\mid Z=0,X,U])}{(1-E[D\mid Z=1,X,U])E[D\mid Z=0,X,U]}:$
\begin{enumerate}
    \item A change in $X$ or $U$ would result in a smaller change in $RD(X,U)$ if the baseline risk was near 0 or 1 compared to near 0.5 (a statement made by \cite{omalley2022discussion});
    \item A change in $X$ or $U$ is equally likely to result in a  change in $OR(X,U)$ if the baseline risk was near 0 or 1 compared to near 0.5 (our interpretation of \cite{omalley2022discussion}'s comment relating to the logistic model);
     \item A change in $X$ or $U$ is equally likely to result in a  change in $RD(X,U)$ if the odds product was near 0 or $\infty$ compared to near 1.
\end{enumerate}
We believe that statements (1) and (2) appearing reasonable is due to the fact that while $OR(X,U)$ is variation independent of $E[D\mid Z=0,X,U],$ $RD(X,U)$ is not. However, contrary to the claim made by \cite{omalley2022discussion}, this does \emph{not} imply $RD(X,U)$ is less likely to be homogeneous. For example, statement (3) appears as reasonable as statement (2). Although the baseline risk is a nuisance parameter that is more natural to humans,  there is no reason to believe that nature also prefers the baseline risk over the odds product as a nuisance parameter. We refer interested readers to \cite{wang2022homogeneity} for detailed discussions and alternative nuisance parameters.

 \subsection{Further comments on Assumption 5 in Wang et al. (2022)}
	
	
We point out that as noted after  \citet[][Assumption 5]{wang2022instrumental}, when the instrument Z is randomized,  Assumption 5 holds as long as all unmeasured confounders for the effect of $D$ on $Y$ do not predict compliance type.  Furthermore, Proposition 5 can be used if Assumption 5 is violated.
	
	We agree with \cite{omalley2022discussion}'s suggestion that a sensitivity analysis assessing the robustness is needed, and leave this as future work.

\section{Identification of related causal parameters}
\label{sec:identification}

\subsection{Identification of the cumulative distributions}

\cite{frandsen2022discussion} suggests considering the cumulative distributions of the treated and untreated potential durations, defined as
    $F_1(t) = Pr(T(1)\leq t),     F_0(t) = Pr(T(0)\leq t).$
These parameters are indeed identifiable under the no additive $U-Z$ interaction in the treatment model, i.e. Assumption  5 in \cite{wang2022instrumental}; see \citet[][Theorem 1]{cui2021instrumental} for a more general result that also applies to the setting with time-varying treatments.

To provide intuition, recall that \cite{wang2018bounded} showed that $E\{Y(1)-Y(0)\}$ is identifiable under the assumption of no additive $U-Z$ interaction in the treatment model. If we let $Y = DI(T\leq t),$ then the results in \cite{wang2018bounded} show that in the absence of censoring, $E\{Y(1)-Y(0)\} = E\{I(T(1)\leq t)\} = F_1(t)$ is identifiable. Similarly one can show identifiability of $F_0(t)$ by letting $Y = -(1-D)I(T\leq t).$

\subsection{Identification of the complier hazard ratio}

\cite{frandsen2022discussion} also suggests considering the effects for the compliers. Before discussing this in the survival context, we first review related results for  the uncensored outcome. In particular, let $\delta^Y(X) = E(Y\mid Z=1,X)-E(Y\mid Z=0,X), \delta^D(X) = E(D\mid Z=1,X)-E(D\mid Z=0,X)$. \cite{wang2018bounded} pointed out that under various sets of assumptions, conditional on measured covariates $X$, the average treatment effect $ATE(X) = E[Y(1)-Y(0)\mid X],$ the local average treatment effect $LATE(X) = E[Y(1)-Y(0)\mid D(1)>D(0),X]$, and the effect of treatment on the treated $ETT(X) = E[Y(1)-Y(0)\mid D=1,X]$ conincide with each other. However, the marginal effects differ as the distribution of $X$ may differ in the entire population, the compliers, and the treated subgroup. Specifically, while the conditional Wald estimand  $\delta^Y(X)/\delta^D(X)$ identifies $ATE(X), LATE(X)$ and $ETT(X)$ under their respective identifying conditions, 
\begin{flalign*}
ATE &= E_X ATE(X) = E_X \dfrac{\delta^Y(X)}{\delta^D(X)} \quad \text{assuming no unmeasured common effect modifier}; \\
LATE &= E_{X\mid D(1)>D(0)} LATE(X) = \dfrac{E_X \delta^Y(X)}{E_X \delta^D(X)} \quad \text{assuming monotonicity}; \\
ETT &= E_{X\mid D=1} ETT(X) = E_{X\mid D=1} \dfrac{\delta^Y(X)}{\delta^D(X)}  \quad \text{assuming no current treatment value interaction} .
\end{flalign*}

In the survival context,
\cite{omalley2022discussion} note that in the case of no measured covariates, the estimating equation (9) in \cite{wang2022instrumental} can also be used to estimate the complier hazard ratio. Nevertheless, we do not recommend using our estimator to estimate the marginal complier hazard ratio when there are observed covariates, as (1) the covariate distribution may differ in the complier stratum and the entire population; (2) the hazard ratio is not collapsible, so that in general, the marginal hazard ratio  \emph{cannot} be expressed as weighted averages of conditional or stratum-specific hazard ratios.

\section{A specification test for the proportional hazard assumption in the IV context}
\label{sec:specification}

\cite{frandsen2022discussion} calls for a specification test for the proportional hazards assumption in the IV context. Recall that in \cite{wang2022instrumental}, we show that $\widehat{\psi}$ is a consistent estimator of $\psi_0$, and further that
 $\mathbb{P}_n H\left(\psi_0, \theta_0\right)=\mathbb{P}_n H^c\left(\tau;\psi_0, \theta_0\right)+o_p(1 / \sqrt{n})$, where we have included a $\tau$, denoting the end of study time-point, in the definition of $H^c\left(\tau;\psi_0, \theta_0\right)$. Thus,
$$
H^c\left(\tau;\psi_0, \theta_0\right)=\int_0^{\tau}\left[\left\{\gamma_1(y)-\gamma_2^{m_0}(y)\right\} D-(1-D) \gamma_2^{m_0}(y)\right] \omega_0(Z, X)\left\{e^{-\psi D} d N(y)-R(y) d \Lambda_0(y)\right\}
$$
with $R(y)=I(Y \geqslant y)$ and $\Lambda_0(y)=\int_0^y \lambda_0(s) d s$. The   $H_i^c\left(\tau;\psi_0, \theta_0\right)$ 's are zero-mean terms that are independent and identically distributed.

We may employ the idea developed in  \cite{lin1993checking}  to check the proportional hazards
assumption. Consider the process
$$
H^n(t;\hat{\psi}, \widehat{\theta})=\mathbb{P}_n\int_0^t\left[\left\{\widehat{\gamma}_1(y)-\widehat{\gamma}_2^{m_0}(y)\right\} D-(1-D) \widehat{\gamma}_2^{m_0}(y)\right] \widehat{\omega}_0(Z, X) e^{-\hat{\psi} D} d N(y).
$$
Assume that the marginal structural Cox model (i.e. equation (1) in \cite{wang2022instrumental}) and the nuisance models are correctly specified. Then one can
write the test-process $n^{1/2}H^n(t;\hat{\psi}, \widehat{\theta})$ as a sum of $n$ zero-mean independent and identically distributed terms which allows us to make draws from the limit distribution
of the test-process  under the null using the resampling procedure outlined in  \cite{lin1993checking}.
 A formal test may be developed using for instance $$\mathrm{TST} \equiv \sup _{t \leqslant \tau}\left|n^{1 / 2} H^n(t;\hat{\psi}, \widehat{\theta})\right|$$ as the test statistic. The limit distribution for TST under the null can be approximated using the alluded resampling procedure.

\section[Comment on Baer et al. (2022)’s multiply robust estimator]{Comment on \cite{baer2022discussion}'s multiply robust estimator}

\label{sec:mr}

\cite{baer2022discussion} focus on the statistical estimand in Theorem 2 in \cite{wang2022instrumental}, and derive the efficient influence function in a nonparametric model for this estimand. Although this influence function may no longer be efficient in the semiparametric Cox marginal structural model, it is still a valid influence function. Motivated by this, \cite{baer2022discussion} develop an augmented variant of \cite{wang2022instrumental}'s estimator, and show that it is multiply robust. This mirrors the development of a multiply robust estimator in the uncensored context by \cite{wang2018bounded}. We congratulate the authors on an intriguing and thoughtful development. 

\cite{baer2022discussion}'s results provide insights into a quest by \cite{frandsen2022discussion}, who asks for an analysis of the efficiency gains from imposing the proportional hazard assumption.  Although the plug-in estimator by \cite{wang2022instrumental} is not necessarily the most efficient one within the semiparametric family imposing the proportional hazard assumption, and the augmented estimator by \cite{baer2022discussion} is not necessarily the most efficient one in the nonparametric family either due to omitting the last two terms in the nonparametric efficient influence function in Proposition 1, their comparison still shed light into the efficiency gain due to the proportional hazards assumption. For example, under correct model specifications, \citet[][Table 1]{baer2022discussion} shows that the standard error for the plugin estimator is 0.23, while that for the augmented estimator is 0.26. The difference in terms of efficiency can be interpreted as the price paid for the multiple robustness enjoyed by the augmented estimator but not the plugin estimator.

 \section{Other comments}
 \label{sec:others}

 \subsection{The use of negative weights}
 
 \cite{omalley2022discussion} pointed out that the weights in \cite{wang2022instrumental}, $\omega(Z,D) = (2Z-1)h(D)/f(Z\mid X)$ can be negative. We note that if $h(D)=1,$ then this weight function is widely used in other contexts in the  causal inference literature, such as the g-estimation and inverse probability weighting. For example, consider the widely used inverse probability weighting estimator for the average treatment effect \citep{hernan2020causal}:
 $$
    \hat{\Delta}_{IPW}  = \mathbb{P}_n \left\{ \dfrac{ZY}{e(X)} - \dfrac{(1-Z)Y}{1-e(X)}\right\} = \mathbb{P}_n \dfrac{2Z-1}{f(Z\mid X)} Y,
 $$
 where $e(X) = P(Z=1\mid X)$ is the propensity score, and $\mathbb{P}_n$ is the empirical mean operator. 
 
 \subsection{Marginal versus conditional survival models}
 
 \cite{omalley2022discussion} raise an interesting question of whether it is better to have a simple expression for the marginal model $S_d^T(t)$ versus the conditional model $S_d^T(t\mid X=x,U=u)$. We first state the obvious: for a given $U$, neither model is stronger or weaker, and they are in general not nested. We nevertheless prefer to specify a marginal model for two reasons. First, it is easier to check the assumptions, such as the proportional hazards assumption, for the marginal model.  Second, in general, there can be many sets of unmeasured confounders $U$ that satisfy Assumptions 1, 3, 5 in \cite{wang2022instrumental}. Due to non-collapsibility of hazard ratio, proportional hazards for one set of $U$ may not be compatible with proportional hazards for a different set of $U.$ Even if they are compatible in some special settings, the true parameter values in a conditional model also varies depending on the choice of $U$.  Hence it is hard to define a causal model condition on unmeasured confounders without observing/specifying them.

 \subsection[The simulation studies in Wang et al. (2022)]{The simulation studies in \cite{wang2022instrumental}}
 
 \cite{omalley2022discussion} conduct additional simulation studies showing that when there are no measured covariates, \cite{mackenzie2014using}'s method is comparable to \cite{wang2022instrumental}'s.  We note, however, that a major limitation of \cite{mackenzie2014using}'s approach is that it is not designed to be used with covariates, as they assume the instrument to be not only valid for the treatment, but also for all the covariates.  The simulations  in \cite{wang2022instrumental} were designed to illustrate this point: if the instrument is only valid conditional on a set of covariates, then \cite{mackenzie2014using}'s estimator may not perform well. 
 
 We further comment on the interpretation of the bias in both \citet[][Table 1]{wang2022instrumental} and \citet[][Table 2]{omalley2022discussion}. The largest bias of the crude Cox model estimate is 0.057, with a standard error of 0.0029. As the bias is much larger than the standard error, we concluded that the crude estimator is severely biased. We disagree with \cite{omalley2022discussion}'s presentation of this bias as 5.7\%, which obscures the fact that this number represents the \emph{absolute value} of the bias. A similar comment applies to the statement that under \cite{omalley2022discussion}'s simulation settings, ``the Cox-crude results having up to nearly 100\% bias.''

	\thispagestyle{empty}
	\bibliographystyle{apalike}
	\bibliography{causal}

\begin{thebibliography}{}

\bibitem[Baer et~al., 2022]{baer2022discussion}
Baer, B.~R., Strawderman, R.~L., and Ertefaie, A. (2022).
\newblock Discussion on ``{I}nstrumental variable estimation of causal hazard
  ratio''.
\newblock {\em Biometrics}.

\bibitem[Cui et~al., 2021]{cui2021instrumental}
Cui, Y., Michael, H., Tanser, F., and Tchetgen~Tchetgen, E. (2021).
\newblock Instrumental variable estimation of the marginal structural cox model
  for time-varying treatments.
\newblock {\em Biometrika}.

\bibitem[Frandsen, 2022]{frandsen2022discussion}
Frandsen, B.~R. (2022).
\newblock Discussion on ``{I}nstrumental variable estimation of causal hazard
  ratio''.
\newblock {\em Biometrics}.

\bibitem[Hern{\'a}n and Robins, 2020]{hernan2020causal}
Hern{\'a}n, M.~A. and Robins, J.~M. (2020).
\newblock {\em Causal Inference: What If}.
\newblock Boca Raton: Chapman \& Hall/CRC.

\bibitem[Lin et~al., 1993]{lin1993checking}
Lin, D.~Y., Wei, L.-J., and Ying, Z. (1993).
\newblock Checking the {C}ox model with cumulative sums of martingale-based
  residuals.
\newblock {\em Biometrika}, 80(3):557--572.

\bibitem[MacKenzie et~al., 2014]{mackenzie2014using}
MacKenzie, T.~A., Tosteson, T.~D., Morden, N.~E., Stukel, T.~A., and O'Malley,
  A.~J. (2014).
\newblock Using instrumental variables to estimate a {C}ox's proportional
  hazards regression subject to additive confounding.
\newblock {\em Health Services and Outcomes Research Methodology},
  14(1-2):54--68.

\bibitem[O'Malley et~al., 2022]{omalley2022discussion}
O'Malley, A.~J., Pablo, M.-C., and MacKenzie, T.~A. (2022).
\newblock Discussion on ``{I}nstrumental variable estimation of causal hazard
  ratio''.
\newblock {\em Biometrics}.

\bibitem[Richardson et~al., 2017]{richardson2017modeling}
Richardson, T.~S., Robins, J.~M., and Wang, L. (2017).
\newblock On modeling and estimation for the relative risk and risk difference.
\newblock {\em Journal of the American Statistical Association},
  112(519):1121--1130.

\bibitem[Wang, 2022]{wang2022homogeneity}
Wang, L. (2022).
\newblock On the homogeneity of measures for binary associations.
\newblock {\em arXiv preprint arXiv:2210.05179}.

\bibitem[Wang et~al., 2022]{wang2022instrumental}
Wang, L., Tchetgen, E.~T., Martinussen, T., and Vansteelandt, S. (2022).
\newblock Instrumental variable estimation of the causal hazard ratio.
\newblock {\em Biometrics}.

\bibitem[Wang and Tchetgen~Tchetgen, 2018]{wang2018bounded}
Wang, L. and Tchetgen~Tchetgen, E. (2018).
\newblock Bounded, efficient and multiply robust estimation of average
  treatment effects using instrumental variables.
\newblock {\em Journal of the Royal Statistical Society: Series B (Statistical
  Methodology)}, 80:531--550.

\end{thebibliography}

\end{document}